\documentclass[%
 reprint,
superscriptaddress,
 amsmath,amssymb,
 aps,
 pra,
longbibliography
]{revtex4-2}

\usepackage{bm}
\usepackage{graphicx}
\usepackage{xcolor}

\newcommand{\bb}{\mathbf}

\newcommand{\ee}{\mathrm{e}}
\newcommand{\dd}{\mathrm{d}}

\begin{document}

\title{Optical force density and surface displacements in transparent dielectrics due to non-ionizing sub-picosecond laser excitation}



\author{B. Anghinoni}
\email{brunoanghinoni@gmail.com}
\affiliation{Department of Physics, Universidade Estadual de Maring\'a, Maring\'a, PR 87020-900, Brazil}
\author{L. C. Malacarne}
\affiliation{Department of Physics, Universidade Estadual de Maring\'a, Maring\'a, PR 87020-900, Brazil}
\author{G. V. B. Lukasievicz}
\affiliation{Department of Physics, Universidade Tecnológica Federal do Paran\'a, Medianeira, PR 85884-000, Brazil}
\affiliation{Institute of Chemical Technologies and Analytics, Technische Universität Wien, Getreidemarkt 9/164, 1060 Vienna, Austria}
\author{B. Lendl}
\affiliation{Institute of Chemical Technologies and Analytics, Technische Universität Wien, Getreidemarkt 9/164, 1060 Vienna, Austria}
\author{M. L. Baesso}
\affiliation{Department of Physics, Universidade Estadual de Maring\'a, Maring\'a, PR 87020-900, Brazil}
\author{N. G. C. Astrath}
\email{ngcastrath@uem.br}
\affiliation{Department of Physics, Universidade Estadual de Maring\'a, Maring\'a, PR 87020-900, Brazil}
\affiliation{Institute of Chemical Technologies and Analytics, Technische Universität Wien, Getreidemarkt 9/164, 1060 Vienna, Austria}


\begin{abstract}
The optical force density acting in transparent dielectric media due to short laser excitation is theoretically analyzed. For typical laser pulses with picosecond duration, the momentum component of the optical force becomes of the same order of magnitude as the stress component, enhancing the overall optomechanical effects. Simulations of the optically-induced surface displacements in fused silica glass are also presented. With careful choice of realistic simulation parameters, dispersive and nonlinear effects were shown to be suppressed and displacements of 50 pm were found for single-pulse excitation with 100~fs duration, where the momentum component of the force is dominant. A possibility of measuring these displacements with piezo-electric detection is also discussed, providing a way to attain spatiotemporal characterization of the optical momentum force in continuum media. Such a description would significantly advance our current knowledge on the behavior of optical forces, potentially contributing to more versatile optomechanical applications, while also clarifying the ongoing Abraham-Minkowski fundamental problem on the momentum transferred by light.  
\end{abstract}

\maketitle

\section{Introduction}\label{sec:intro}

Precise knowledge of the optical forces acting in matter under laser illumination is integral to modern optomechanics, finding applications, for example, in the design of photonic devices~\cite{Wiederhecker2009,Partanen2022b}, waveguides~\cite{Rakich2010,Pernice2009,Wang2016}, optofluidics~\cite{Monat2007,Psaltis2006} and optical manipulation of biological systems~\cite{Brevik2023,Torbati2022}. From classical field theory, it is known that forces are described through a four-continuity equation associated to a stress-energy-momentum (SEM) tensor~\cite{LL_fields}. In this context, forces in continuum matter originate either from stress effects or from momentum transfer and, for optical forces, the momentum transfer occurs from light to the medium. Describing such transfer, however, is far from trivial -- in fact, it represents a long-standing dilemma in physics, which started more than a hundred years ago and is currently known as the Abraham-Minkowski (A-M) problem~\cite{Abraham,Minkowski}. This fundamental process has proven to be very elusive, where only few experimental works were capable of assessing light's momentum~\cite{Jones1954,Jones1978,Gibson1980,Strait2019,Campbell2005}. On the other hand, theoretical works are abundant and present very rich discussions with distinct formulations and interpretations for the A-M problem~\cite{Brevik1979,Kemp2011,Pfeifer2007,EL,Anghinoni2022,Anghinoni2023,Anghinoni2024,MP1,Milonni2010,Barnett2010,Correa2020,Silveirinha2017,Bethune-Waddell2015}.

The propagation of light in continuum material media inevitably generates stress effects, whose dynamics have been investigated in earlier literature~\cite{Gordon1973,Anghinoni2024,Astrath2024,Partanen2023}. Distinguishing these optically-induced stress effects from those stemming from optical momentum transfer is crucial for a correct interpretation of the light-matter interactions and the A-M problem~\cite{Brevik2018c}. Indeed, attaining such distinction experimentally is a formidable task as bulk stress forces do not affect the center of mass movement of a body~\cite{Brevik1979,Kemp2017} -- nevertheless, the spatiotemporal distribution of such stress forces has been successfully addressed in transparent liquids~\cite{Astrath2022,Astrath2023,Astrath2024}. The spatiotemporal distribution of the momentum part of the bulk optical force, on its turn, is typically very small in magnitude and has never been observed. However, contrary to the stress counterpart, once this momentum force distribution is integrated over the sample's volume we obtain a contribution to the total optical force acting on the sample that, in some situations, is capable of producing measurable effects on the center of mass dynamics of the body. Such dynamics have been studied experimentally both under low-frequency~\cite{Walker1975,Rikken2012} and optical excitations~\cite{Jones1954,Jones1978}. 

As discussed in detail the next section, to enhance the optical momentum transfer and render its effects more appreciable, we generally need excitations with relatively fast modulations in time. 
Ideally, high beam intensities should also be applied as the optical force density scales linearly with intensity. These two facts together bring about the potential necessity of a full nonlinear and dispersive description of the beam propagation, which would make the problem much more complicated and hinder any effort to isolate the optomechanical effects stemming from momentum transfer. Nevertheless, we will show that by choosing suitable sample material and geometry along with some realistic beam parameters, it is possible to have a propagating beam whose intensity is large enough to produce measurable optically-induced surface deformations on the sample while not being significantly affected by nonlinear phenomena or temporal dispersion during propagation. This behavior will greatly simplify the theoretical analysis of the deformations, from which the contribution of the optical momentum force component can be clearly characterized in space and time.

This work is organized as follows. In Sec.~\ref{sec:f} we theoretically describe the optical force density acting in transparent dielectric media due to the propagation of a short laser pulse. It is seen that the momentum part of the optical force acquires the same order of magnitude as the stress part for typical pulsed beams of about 1 ps duration. In Sec.~\ref{sec:pulse}, we show that the potentially complicating dispersion and nonlinear effects can be neglected in the propagation of a 100~fs single-pulse gaussian beam with 0.5~TW/cm$^2$ intensity and 100~$\mu$m radius by choosing a fused silica glass sample with millimetric size and an appropriate optical setup. In Sec.~\ref{sec:num} we simulate the optical forces and the related surface deformations of this sample, showing that displacements of approximately 50 pm can be generated solely by the momentum part of the optical force. We also propose and discuss the possibility of measuring the aforementioned displacements by employing piezo-electric detection, aiming to characterize the optical momentum transfer in space and time. In Sec.~\ref{sec:disc}, an analysis of the earlier specialized literature is provided along with comparisons to our current approach. At last, conclusions are drawn in Sec.~\ref{sec:conc}.
\section{Optical force density in transparent media}\label{sec:f}

The conservation of momentum and energy can be described by a four-continuity equation of the form~\cite{LL_fields} $\partial_{\nu}\mathcal{T}^{\mu\nu}=-f^{\mu}$, where $\mu,\nu=0,1,2,3$, $x^{\nu}=(ct,x,y,z)$, $\partial_{\nu}=\partial/\partial x^{\nu}$, $\mathcal{T}^{\mu\nu}$ is the SEM tensor, $f^{\mu}$ is the four-force density and $c$ is the speed of light in vacuum. Here, summation over repeated indices is implied. Focusing on the stress-momentum sector ($\mathcal{T}^{i\nu}$ with $i=1,2,3$), we obtain the momentum conservation equation as
\begin{equation}\label{eq:T}
    \frac{\partial \bb g}{\partial t}+\bm\nabla\cdot\overleftrightarrow{\bb T} = -\bb f,
\end{equation}
where $\bb g$ is the momentum density, $\overleftrightarrow{\bb T}$ the stress tensor and $\bb f$ the force density. Naturally, the addition of the appropriate material's SEM tensor would in our case completely delineate the closed system of interest, building a total SEM tensor whose four-divergence is necessarily zero. For a more detailed discussion on the SEM tensor of matter plus electromagnetic fields in the context of optical forces, see e.g. Refs.~\cite{Anghinoni2024,Partanen2023,MP6}.

Under the dipolar approximation, we adopt the optical force density in linear, homogeneous, non-dispersive, lossless and non-magnetic bulk media as~\cite{Anghinoni2023,Astrath2024}
\begin{equation}\label{eq:f}
    \bb f = \frac{1}{2}\bm\nabla(\bb P\cdot \bb E)+\frac{n^2-1}{c^2}\frac{\partial}{\partial t}(\bb E \times \bb H).
\end{equation}
In the above equation, $\bb E$, $\bb P$ and $\bb H$ are the electric, polarization and magnetic fields, respectively, and $n$ is the refractive index of the medium. Comparing Eqs.~(\ref{eq:T}) and~(\ref{eq:f}), we can identify $\overleftrightarrow{\bb T}\to(\bb P \cdot \bb E)\overleftrightarrow{\bb I}/2$ (where $\overleftrightarrow{\bb I}$ denotes the unit dyadic) and $\bb g \to c^{-2}(n^2-1)\bb E \times \bb H$. Therefore, the first term of Eq.~(\ref{eq:f}) corresponds to a stress force, which is known as electrostriction effect, while the last term corresponds to the momentum part of the force and is known as Abraham force density. Specifically, this momentum component averages to zero in applications that employ harmonic fields, such as in typical optical trapping setups -- therefore, it is, in principle, only relevant when non-harmonic fields are present. 

\subsection{Optical force density with pulsed gaussian beams}\label{sec:f_gauss}

For tightly focused quasi-monochromatic gaussian beams, the optical force density $\bb f$ can be obtained semi-analytically through the Angular Spectrum method~\cite{ASRWolf}, where the numerical procedures involved are reduced to calculating simple finite, one-dimensional integrals~\cite{Anghinoni2023b}. In typical setups, however, the deviation of the focused beam from the paraxial approximation is very small -- thus, it is possible to adopt the known analytical solutions of the Helmholtz wave equation in the paraxial regime~\cite{Pedrotti1993}. For a linearly polarized gaussian beam in fundamental mode propagating in the $z$-direction we have (see Appendix~\ref{app:f} for the calculation)
\begin{equation}\label{eq:f_I}
    \bb f(r,z,t)\!=\!-\frac{8}{c}\!\left[\frac{r}{w_0^2}\hat{\bb r}\!+\!\frac{2(t\!-\!\xi)}{c\tau^2} \hat{\bb z}\right]\!\frac{(n\!-\!1)}{(n\!+\!1)}I(r,z,t),
\end{equation}
where $w_0$ is the average beam waist inside the medium, $\tau$ is the temporal beam width, $\xi$ is the time of maximum irradiance and $I$ is the beam intensity as measured in air. Also, we assumed a normal beam incidence from air to the medium. The first term within the brackets corresponds to the electrostriction force density and the second term corresponds to the Abraham force density. 


We can see in Eq.~(\ref{eq:f_I}) that the maximum magnitude of the electrostriction contribution occurs for $r = w_0/\sqrt 2$ and $t=\xi$, reaching a value of 
$K/w_0$, 
where $K=8\varepsilon_0(n-1) E_0^2/[\sqrt 2 (n+1)\exp(1)]$ and $E_0$ is the beam amplitude in air. For the Abraham force density, $r = 0$ and $t=\xi\pm\tau/\sqrt 2$ generate maximum magnitude, yielding a value of $2 K/c\tau$. Notice that when one contribution is at maximum magnitude with respect to time, the other one is zero, in agreement with earlier literature~\cite{Tam1986}. The relative peak magnitude of the Abraham term to the electrostriction term is, therefore, $2 w_0/c \tau$. If we take the realistic value of $w_0 = 100\,\mu$m, this ratio of force densities becomes of order $10^{-12}\tau^{-1}$. 
Thus, observations of the optomechanical effects of the Abraham force density as the dominant mechanism should be possible with typical gaussian beams with $\tau \lesssim 1$~ps. As stated earlier, high beam intensities are also desirable as the optical force density would be larger.
In principle, these two features would require a treatment that includes nonlinear and dispersion effects, rendering Eq.~(\ref{eq:f_I}) invalid. Additionally, potential optically-induced breakdown and damage mechanisms must also be accounted for. Nevertheless, we will show in Sec.~\ref{sec:pulse} that by choosing a fused silica glass sample with 
millimetric size and some appropriate beam parameters, it is possible to overcome these limitations.

\section{Optical pulse propagation}\label{sec:pulse}
We consider an incident pulsed gaussian beam comprising a single pulse propagating inside a disk-shaped fused silica glass sample~\cite{Moore2022}, having thickness $L=1$~mm and radius $R=2.5$~cm. The beam wavelength in air is $\lambda_0=800$~nm, with additional beam parameters $\tau = 100$~fs, $\xi = 0.35$~ps, beam energy $Q=1$~mJ and $w_0=100$~$\mu$m. These values correspond to a peak beam intensity of approximately $0.5$~TW/cm$^2$ (see Appendix~\ref{app:f}). The propagation of this beam is described by the generalized nonlinear Sch\"{o}dinger equation (GNLSE)~\cite{Boyd2020,Brabec1997}, which is a very versatile equation to address beam dynamics in nonlinear and dispersive media. In what follows, many considerations are made to properly delineate the dominant physical processes and define the specific form of the GNLSE to be employed.

With a beam intensity below the TW/cm$^2$ order, we can use the perturbative nonlinear optics regime, where effects on the electric polarization are considered to third order in the electric field~\cite{Brabec2000}. Specifically, third harmonic generation is a process that requires phase matching and hence is also not important in our analysis. Nonlinear interactions in our system thus comprise only the optical Kerr effect and the stimulated Raman scattering. In our medium, the Kerr effect leads to beam self-focusing, where the transverse beam profile gets tighter with propagation and, consequently, beam intensity quickly increases~\cite{Kelley1965}. Such behavior is highly undesired as the beam dynamics is complicated by the introduction of a transversal variation and also the occurrence of light-induced damage to the sample becomes much more likely. Fortunately, this issue can be circumvented by choosing a suitable radius of curvature for the incident beam in order to compensate for the effects of self-focusing. This procedure will be discussed in detail later in this section and in the next section -- for now, we will just assume that no beam self-focusing occurs within our sample. With no significant overall beam diffraction, the space-time coupling term for ultrashort beams can also be neglected~\cite{Gaeta2000}.

Fused silica glass has a wide band gap of about 9~eV~\cite{Astasauskas2020,Vella2011}, which requires six photons at 800 nm to be absorbed (about 1.55~eV per photon) for an electronic transition to occur. Besides multi-photon ionization,
tunneling ionization can also occur, according to the well-known Keldysh model~\cite{Keldysh1965}. Nevertheless, for peak intensities of $1$~TW/cm$^2$ (i.e., double our value) at 800 nm, the total photoionization rate in fused silica is of order $10^{25}$~cm$^{-3}$s$^{-1}$~\cite{Lenzer2008}, which, when multiplied by $\tau$, gives us an estimate for the free electronic density as $10^{12}$~cm$^{-3}$. This value is way below the threshold for plasma formation, which at optical frequencies requires a density of order $10^{21}$~cm$^{-3}$. Moreover, detailed calculations of collision rates showed that pulses with 200~fs and lower are too short for electron avalanche to develop~\cite{Kaiser2000}.
Thus, we can neglect ionization effects and plasma formation, and consequently no dielectric breakdown is assumed to take place.

Another important effect to be considered is the possibility of light-induced damage to the sample. Such damage is undesirable as the whole medium dynamics would be significantly affected, rendering application of the GNLSE invalid. The light-induced damage threshold (LIDT) is often described in terms of beam fluence, which is defined as the integral of beam intensity in time over the beam duration. For gaussian beams, the peak fluence $F$ is given as $F = 2Q/\pi w^2$ (see Appendix~\ref{app:f}), which in our case leads to $F\approx$ 6.4~J$/$cm$^2$. For fused silica glass under pulsed excitation of 100~fs at 800~nm, the LIDT critical fluence is approximately 10~J$/$cm$^2$~\cite{Du1994} -- therefore, our proposed setup falls below the LIDT and suits our theoretical description. In this context, it is worth mentioning that our choice of single-pulse excitation is crucial as the LIDT in dielectrics typically decreases with the number of pulses illuminating the sample~\cite{Lenzer2008}. 

The temporal dispersion of the pulse is described generically by a Taylor expansion of the wavenumber $\beta(\omega)$ around the central angular frequency of the pulse  $\omega_0 = 2\pi c/\lambda_0$ as
 \begin{equation}\label{eq:beta}
     \beta(\omega) = \beta_0 +\sum_{m=1}\frac{(\omega-\omega_0)^m}{m!}\beta_m, 
 \end{equation}
where $\beta_0=2\pi n/\lambda_0$ and $\beta_m=(\partial^{m}\beta/ \partial \omega^{m})_{\omega=\omega_0}$. The values of the $\beta_m$ coefficients can be obtained from the Sellmeier equation for $n(\lambda_0)$~\cite{Weber2003}. In our application, we keep terms to third order in the expansion shown in Eq.~(\ref{eq:beta}). At last, the wavenumber $\beta(\omega)$ is assumed to be real, i.e., the medium is lossless.


The considerations made above allow us to analyze the evolution of the beam shape by solving the GNLSE only in the propagation direction $z$ and time $t$ as follows~\cite{Boyd2020}

\begin{eqnarray}\label{eq:GNLSE}
    \frac{\partial A}{\partial z}\!=\!-\frac{i\beta_2}{2} \frac{\partial^2 A}{\partial T^2}+\frac{\beta_3}{6} \frac{\partial^3 A}{\partial T^3}
    \!+i\gamma\left(\!1+\frac{i}{\omega_0}\frac{\partial}{\partial T}\right)\!P{_\mathrm{nl}},    
\end{eqnarray}
where $P{_\mathrm{nl}}$ denotes the nonlinear source term, given as
\begin{equation}\label{eq:P_nl}
    P{_\mathrm{nl}} = (1-\delta_\mathrm R) A |A|^2+\delta_\mathrm R A \int_0^{\infty}h_\mathrm R(t')|A(z,T-t')|^2\dd t'.
\end{equation}
In Eq.~({\ref{eq:GNLSE}}), the variable $T=t-z/v_\mathrm g$ is the retarded time of a pulse traveling with group velocity $v_\mathrm g=1/\beta_1$, $A(z,T)$ is the slowly-varying envelope function of the electric field, whose magnitude in the original coordinates is $E(x,y,z,t)=A(z,t)G(x,y)\ee^{i(\beta_0 z-\omega_0 t)}$, with $G(x,y)$ denoting the spatial gaussian dependence. Also, $\gamma = \omega_0 n_2/c$, where $n_2=2.73\cdot10^{-16}$~cm$^2$/W is the nonlinear refractive index coefficient, assumed to not depend on frequency~\cite{Milam1976}. In Eq.~(\ref{eq:P_nl}), we have~\cite{Agrawal2019} $\delta_\mathrm R=0.245$, with the integral term representing the stimulated Raman scattering and the remaining term representing the Kerr effect.  Specifically, the Kerr effect corresponds to an instantaneous electronic response, while the Raman scattering corresponds to a delayed nuclear response characterized in fused silica glass by the response function~\cite{Lin2006}
\begin{eqnarray}\label{eq:Raman}
    h_\mathrm R(t) = (1-f_\mathrm b)\frac{(\tau_1^2+\tau_2^2)}{\tau_1 \tau_2^2}\ee^{-\frac{t}{\tau_2}}\sin\left(\frac{t}{\tau_1}\right)\nonumber \\
    +f_\mathrm b \frac{2 \tau_\mathrm b-t}{\tau_\mathrm b ^2}\ee^{-\frac{t}{\tau_\mathrm b}},
\end{eqnarray}
where $f_\mathrm b=0.21$, $\tau_1=12.2$~fs, $\tau_2=32$~fs and $\tau_\mathrm b=96$~fs.

\subsection{Characteristic lengths}

Equation~(\ref{eq:GNLSE}) is analyzed in terms of characteristic lengths that quantify the relative contribution of each physical mechanism. Such lengths correspond to distance scales after which it is expected that the related mechanism becomes appreciable in magnitude and, therefore, plays a significant role in the beam propagation. First, the so-called dispersion length, $L_\mathrm d$, comprises the first and second term on the right-hand side of Eq.~(\ref{eq:GNLSE}), and is estimated through the dominant term as $L_\mathrm d = \tau^2/|\beta_2|\approx 28$~cm, where we used $\beta_2 = 362$~fs$^2$/cm. As $L_\mathrm d \gg L$, dispersion is not important in our system. Next, we deal with the last term of Eq.~(\ref{eq:GNLSE}), which describes the self-steepening effect due to dispersion of group velocity~\cite{Gaeta2000}. Its characteristic length is approximately $L_{\mathrm{ss}}=c\tau/(n_2 I_0)$ (where $I_0$ denotes the peak intensity of the beam). Thus, $L_{\mathrm{ss}}\approx 22 $~cm. Again, this value is much larger than our propagation distance, and hence beam self-steepening can be neglected. The last contribution refers to nonlinear phenomena, given by the before-last term of Eq.~(\ref{eq:GNLSE}). The nonlinear characteristic length is $L_{\mathrm {nl}}=c/(n_2\omega_0I_0)\approx 1.0$~mm. Therefore, we conclude that only nonlinear effects are important to correctly capture our beam propagation dynamics.

As mentioned earlier, the nonlinear effects considered here are the Kerr effect and the stimulated Raman scattering. In our case, the later develops from the spontaneous Raman scattering, depleting the pump energy to build a Stokes wave whose intensity grows roughly as $\exp(g_\mathrm R I z)$~\cite{Headley1995}. Here, $g_\mathrm R$ is the Raman gain coefficient of the medium, which is $2\cdot10^{-11}$~cm/W in fused silica glass~\cite{Stolen1973}. After propagating a distance $z=L$, we thus see that the Stokes wave is amplified by a factor $\exp(2) \sim 7.3$. To estimate the initial intensity of this wave, $I_\mathrm S (0,T)$, we use~\cite{Agrawal2019} $I_\mathrm S (0,T)=\hbar \omega_\mathrm S B/(\pi w_0^2)$, where $\omega_S$ is the central frequency of the Raman gain spectrum, approximately 13.2~THz, $B$ is the bandwidth around this peak, approximately 10~THz~\cite{Stolen1989}, and the Stokes profile is assumed gaussian with same radius as the pump beam. These values yield $I_\mathrm S (0,T)$ of order $10^{-3}$~W/cm$^2$, so that the peak intensity of the amplified wave, $I_\mathrm S (L,T)$, is of order $10^{-2}$~W/cm$^2$. This value is much smaller than the pump beam intensity, showing that the Stokes wave excitation is irrelevant in our system, and the effects of stimulated Raman scattering can then be safely neglected. This justifies \textit{a posteriori} why we have not written the GNLSE for the Stokes wave as well, which would be coupled to the GNLSE for the pump beam seen in Eq.~(\ref{eq:GNLSE})]\cite{Headley1995}.


The last and solely surviving contribution is the Kerr effect. It is known to generate a small position-dependent variation of $n_2 I$ in the refractive index of the medium, whose maximum value in our case is of order $10^{-4}$. This variation causes a cumulative acquisition of phase by the beam during propagation, which leads to the self-phase modulation (SPM) phenomenon. Such phase change, in turn, leads to a spectral broadening of the beam but does not change the beam intensity. As $n_2 > 0$, the Kerr effect also causes beam self-focusing due to the transversal spatial gradient induced in the index of refraction, as stated earlier in this section. The characteristic self-focusing distance is $L_\mathrm{sf}=(2 n w_0^2/\lambda_0 )(P/P_\mathrm{cr}-1)^{-1/2}$, where $P$ denotes beam power and $P_\mathrm{cr}\approx 0.15 \lambda_0^2/n n_2$ is the critical power, which represents the minimum power the beam must have for the self-focusing effect to occur~\cite{Boyd2020}. In our system, $P/P_\mathrm{cr}\approx 32$, which gives $L_\mathrm{sf}\approx 6.5$~mm. As this value is not far from $L$, the beam radius is expected to decrease with propagation, naturally increasing the overall beam intensity with $z$. This behavior can, however, be suppressed with the choice of an appropriate optical setup, as described in the next section.




\subsection{Compensating the self-focusing effect}

If beam diffraction is negligible, the converging half-angle associated with self-focused beam propagation, $\theta_\mathrm{sf}$, can be obtained through Fermat's principle as~\cite{Boyd2020} $\theta_\mathrm{sf}=\sqrt{2 n_2 I_0/n}$. With simple trigonometry, the beam radius at $z=L$, denoted by $w_L$, can be shown to decrease to a value $w_L=(w_0-L \tan\theta_\mathrm{sf})\approx 86$~$\mu$m due to self-focusing -- i.e, a $\Delta w=14$~$\mu$m difference. This effect, however, can be compensated by employing an incident gaussian beam that, besides having a radius of 100~$\mu$m at the sample surface, would diverge when propagating through the sample in a way that $w_L$ would be $w_0+\Delta w=114$~$\mu$m in the absence of self-focusing. In this scenario, the natural diffraction of the beam would approximately cancel the undesired focusing process, keeping the transversal profile of the beam constant.

To find the appropriate optical setup that fulfill the requirements given in last paragraph, consider Fig.~(\ref{fig:setup}). Through ray transfer matrix analysis (see Appendix~\ref{app:ray} for the details), we find that for the realistic parameters $w_\mathrm{c}=1.5$~mm (radius of the incoming collimated beam) and $f=7.5$~cm (lens focal length), the distance $d$ that solves our problem is $d=5$~mm. Accounting for the self-focusing, this specific choice and placement of optical elements is effectively able to produce a beam intensity that maintain its initial transversal profile while propagating through the sample, allowing for a safe application of the GNLSE in terms of only $z$ and $t$, as seen in Eq.~(\ref{eq:GNLSE}). Further confirmation of this self-focusing canceling can be seen through Snell's law of refraction. The incident beam has a divergence angle $\theta_1 = \lambda_0 / \pi w_{\mathrm{min}}$, where $w_{\mathrm{min}}=[w_\mathrm c^{-2}+(\pi w_\mathrm c/\lambda_0 f)^2]^{-1/2}\approx 12.7$~$\mu$m~\cite{Pedrotti1993}. Upon refraction, the new beam divergence angle is $\theta_2 = \sin^{-1}(\sin\theta_1 /n)$, which can be numerically verified to match $\theta_\mathrm{sf}$, as expected. 

\begin{figure}[htb]
    \centering
    \includegraphics[width=1.0\linewidth]{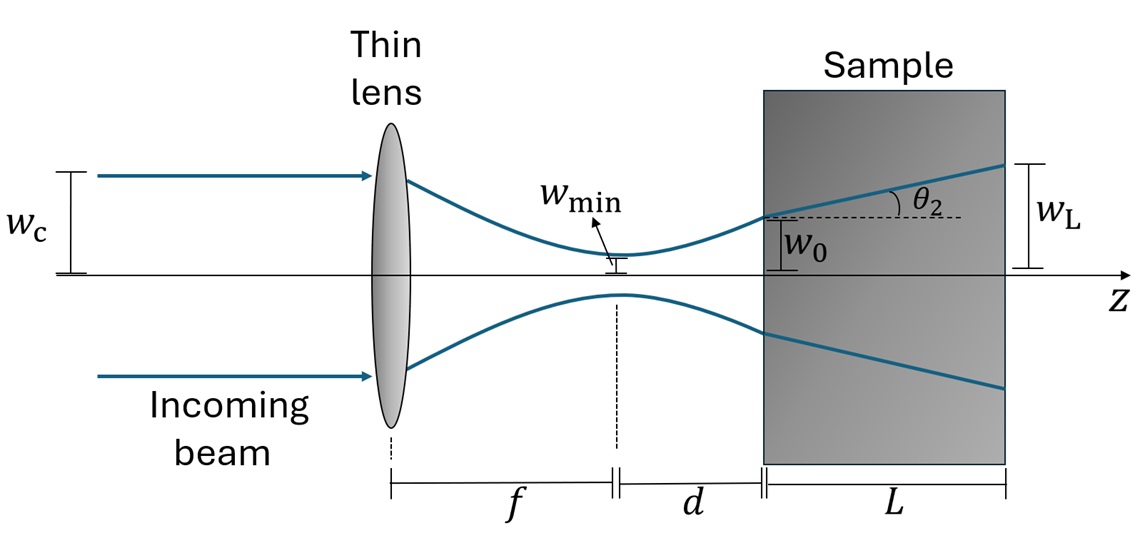}
    \caption{Optical setup to suppress self-focusing. A collimated beam of radius $w_\mathrm c$ is focused by a lens of focal length $f$, with the focal spot placed a distance $d$ of the sample's surface. For our purposes, these three parameters must be such that the beam radii $w_0$ and $w_L$ are $100$~$\mu$m and $114$~$\mu$m, respectively.}
    \label{fig:setup}
\end{figure}

\subsection{Numerical solution of the GNLSE}

\begin{figure}[htb]
    \centering
    \includegraphics[width=1.0\linewidth]{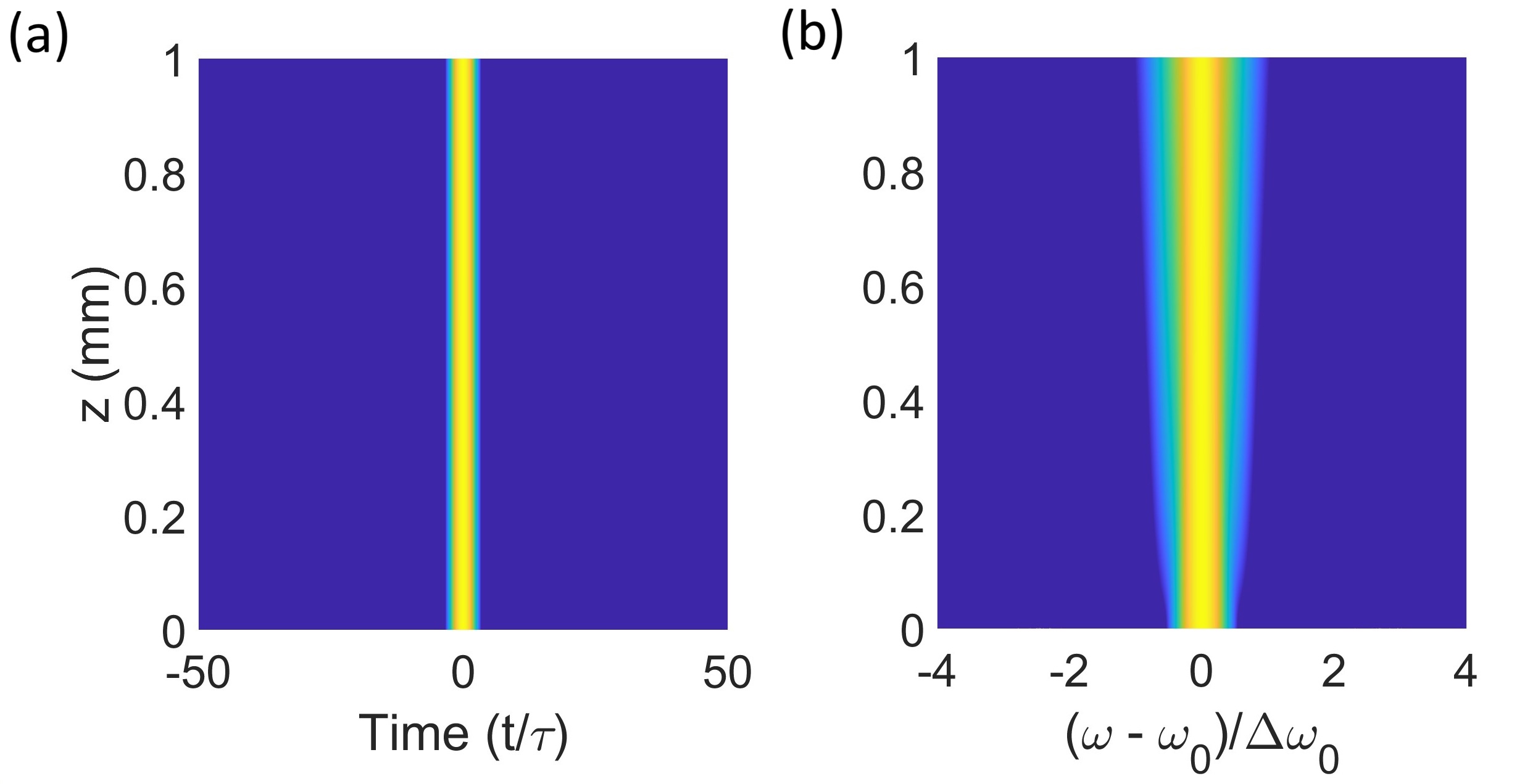}
    \caption{Evolution of the incoming gaussian beam in (a) real space and (b) frequency space, according to the GNLSE in Eq.~(\ref{eq:GNLSE}). Although the SPM effect takes place, beam intensity remains unaltered.}
    \label{fig:spec}
\end{figure}

Through the detailed discussion developed in this section, we expect that in our setup the beam intensity shape does not change significantly when propagating within the sample. Nevertheless, to further validate our analyses and estimates, Eq.~(\ref{eq:GNLSE}) was solved using the split-step Fourier method, whose implementation can be found in Ref.~\cite{Agrawal2019}. The remaining necessary parameter that needs specification is $\beta_3=275$~fs$^3$/cm. Figure~(\ref{fig:spec}) shows the beam dynamics. We see that no relevant changes took place in beam intensity shape (Fig~(\ref{fig:spec}a)), in accordance with the estimates presented in this section. The spectrum, on the other hand, is visibly broadened by SPM, as seen in Fig~(\ref{fig:spec}b). In units of the initial spectral width $\Delta \omega_0$, an upper bound for this spectral broadening for a gaussian beam can be calculated as~\cite{Agrawal2019} $[1+ (2\gamma I_0 L)^2/3\sqrt3 )]^{1/2}$, which in our case yields about 1.4 units, in agreement with Fig~(\ref{fig:spec}b).  

\section{Optically-induced surface displacements}\label{sec:num}

As the beam intensity shape was shown to be unaltered in last section, Eq.~(\ref{eq:f_I}) can be employed to describe the behavior of the optical force. To analyze the mechanical effects generated by this optical force, numerical simulations of the elastic surface dynamics of the fused silica glass sample were carried out. 

\subsection{Optical force density}
To better visualize the optical force density, its two components were numerically calculated for the same pulsed gaussian beam adopted in the last section. 
 
\begin{figure}[htb]
    \includegraphics[width=1.0\linewidth]{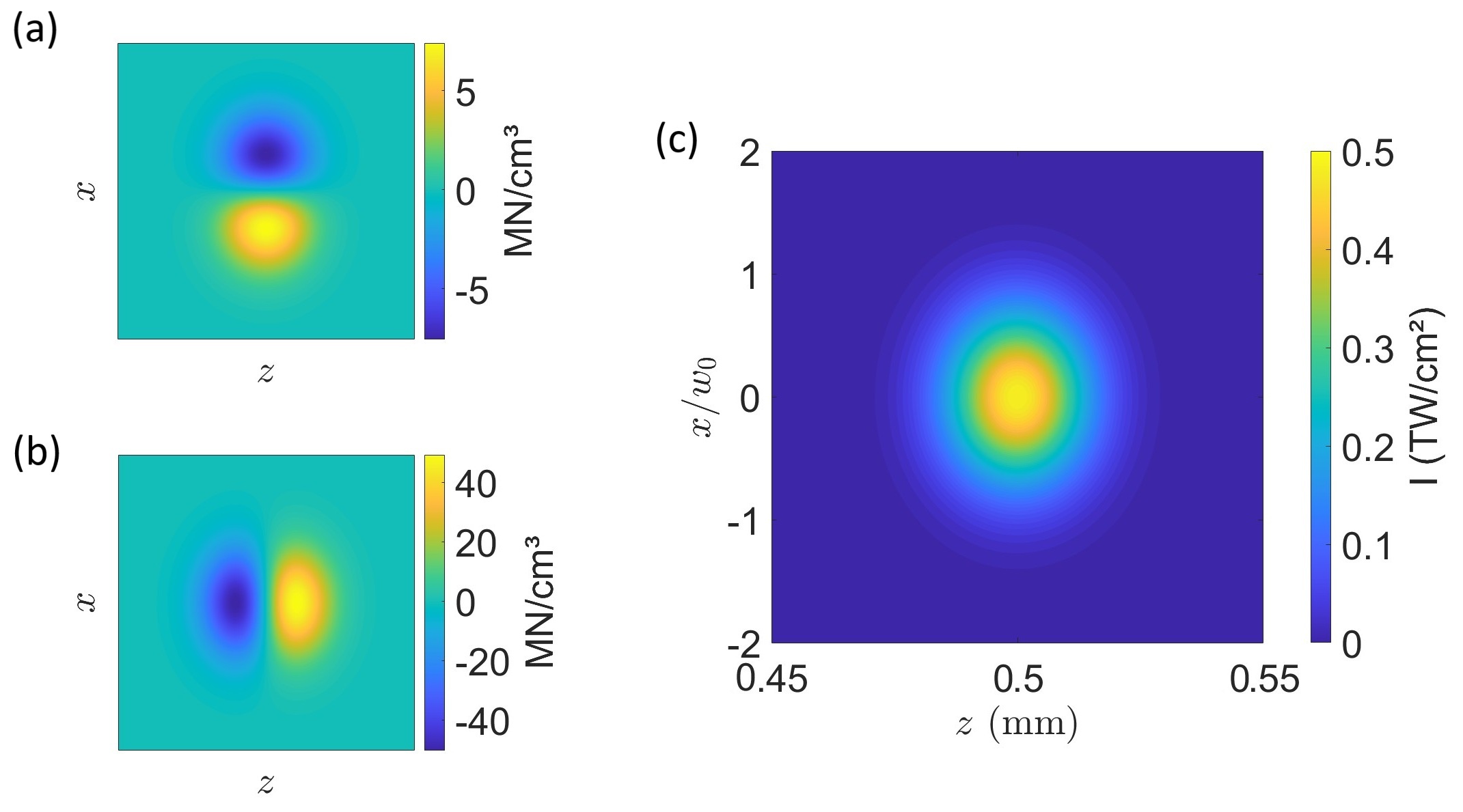}
    \caption{(a) $x$-component of the electrostriction force density. (b) Abraham force density. (c) Beam intensity. All figures refer to the $y=0$ plane.}
    \label{fig:If}
\end{figure}

In Figs.~(\ref{fig:If}a) and~(\ref{fig:If}b) we see the electrostriction and Abraham force densities, respectively, at the $y=0$ plane and when the pulse is at the center of the sample in longitudinal direction ($z$). We also see the beam intensity in Fig.~(\ref{fig:If}c). The two force density components exhibit the expected behavior of symmetric positive and negative values in a lossless medium. Besides, the Abraham contribution is one order of magnitude larger than the electrostriction contribution, in accordance with a short excitation with $\tau = 100$~fs. This situation is different, for example, from the opto-elastic simulations reported in Ref.~\cite{Partanen2023}, where laser pulses with $\tau \sim 100$~ps were considered and, consequently, the electrostriction was seen to completely dominate the total force density. Specifically, the Abraham force density is associated to an expansion-compression wave in the direction of beam propagation. The electrostriction force density, on its turn, is seen here as the $x$-component and compresses the medium towards the center of the beam ($x=0$), where the beam intensity is higher -- an effect long-known to occur in condensed matter~\cite{LL}.



\subsection{Surface elastic dynamics}\label{sec:disp}
The displacements $\bb u$ around the equilibrium configuration of elastic solid media are governed by the thermo-elastic equation. If thermal effects can be neglected, such equation reads~\cite{Nowacki1986}
\begin{equation}\label{eq:TE}
    (1-2\nu)\bm \nabla^2 \bb u+\bm\nabla(\bm\nabla\cdot \bb u) = a\left[\rho\frac{\partial^2\bb u}{\partial t^2}+\bb f_\mathrm{b}\right],
\end{equation}
with $a=2(1+\nu)(1-2\nu)/Y$, where $\nu$ is the Poisson ratio, $Y$ is the Young modulus, $\rho$ is the mass density and $\bb f_\mathrm{b}$ the resultant body force density.

Equation~(\ref{eq:TE}) was numerically solved in the COMSOL Multiphysics software using the finite-element method. The body force density $\bb f_\mathrm b$ is solely given by the optical force density from Eq.~(\ref{eq:f_I}). The sample and beam parameters given earlier to calculate the optical force density were maintained -- except for the beam temporal width $\tau$, where two distinct cases were considered, with $10$~ps or $100$~fs. Very importantly, we notice that considering the excitation with $10$~ps here serves merely for illustration of the deformations expected when the two optical force components are close in value, for all the detailed analyses performed last section pertaining to the pulse propagation are specific to the $100$~fs pulse duration.
The physical properties in Eq.~(\ref{eq:TE}) are taken as~\cite{Weber2003} $\rho = 2200$~kg/m$^3$, $\nu = 0.17$ and $Y=70$~GPa. The boundary conditions employed are null shear stresses at $z=0$ and $z=L$, while normal stresses are given by $(1-\varepsilon) |\bb E(r,z=0,t)|^2/2$ at $z=0$ and $(\varepsilon-1) |\bb E(r,z=L,t)|^2/2$ at $z=L$, where $\varepsilon$ denotes the sample's electric permittivity. Also, the sample is assumed to be surrounded by air. These expressions correspond to the known radiation pressure at dielectric interfaces and can be shown to originate from the electrostriction force density~\cite{Anghinoni2023,Astrath2014}.

\begin{figure}[htb]
    \includegraphics[width=1.0\linewidth]{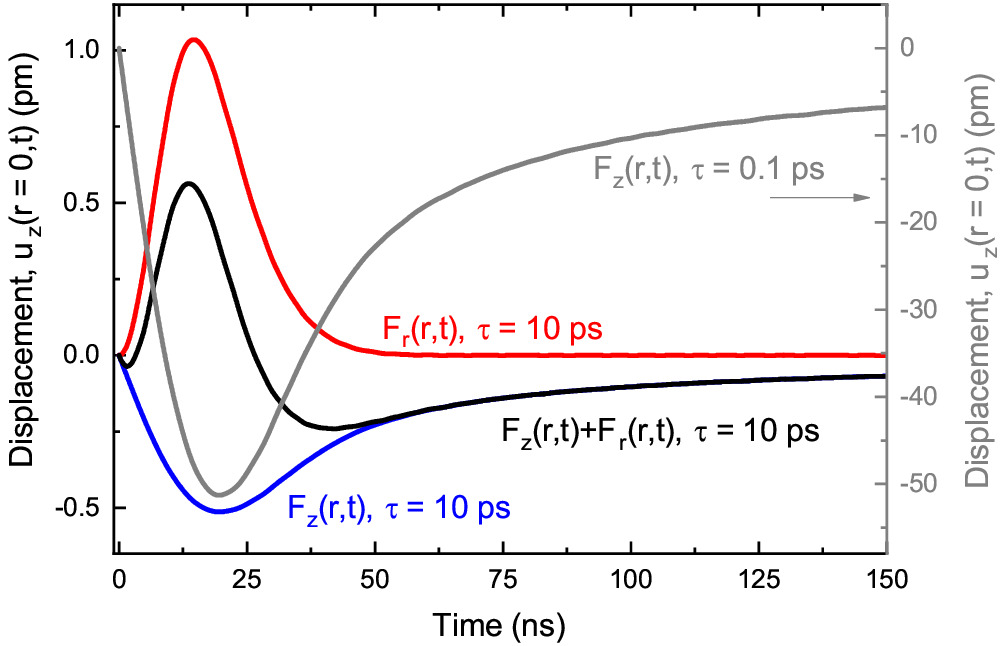}
    \caption{Displacement of the sample surface in the $z$-direction, $u_z$, at $(r,z)=(0,0)$ as a function of time. For $\tau=10$~ps, both optical force components are important (red and blue lines), generating an overall deformation smaller than 1 pm (left axis). For $\tau=100$~fs, the Abraham component (gray line) is solely responsible for the surface deformation, reaching a value of 50 pm (right axis). In this case, the electrostriction contribution is negligible and therefore not shown. Notice the $\tau=10$~ps results are merely illustrative as the beam propagation would be significantly different than previously considered for $\tau=100$~fs.}
    \label{fig:comsol}
\end{figure}

Figure~(\ref{fig:comsol}) shows the displacement in $z$-direction, $u_z$, as a function of time at the surface $z=0$ for the center of the beam ($r=0$) for both considered values of $\tau$. For $\tau=10$~ps, both force components (denoted by $F_r$ for the electrostriction term and $F_z$ for the Abraham term) contribute with picometer-order displacements, seen in red and blue lines. Their contribution present opposite signs, thus generating a sub-picometer resultant displacement, seen in the black line. For $\tau=100$~fs, on the other hand, the optical force
is completely dominated by the Abraham term, causing
a peak surface displacement of approximately 50 pm, shown by the gray line. The electrostriction contribution is this case is negligible and was not shown. Results for the other two sample surfaces are also not shown here, but yield displacements of the same order of magnitude.

\subsection{Possibility of experimental observation}\label{sec:exp}
Studies of the spatio-temporal distribution of the optical force density through the measurement of optically-induced surface deformations in solid samples have been already reported, where displacements of order 1 pm could be observed through piezo-electric surface detection~\cite{Pozar2018,Pozar2013}. Specifically, in Ref.~\cite{Pozar2018} a system similar to the one adopted here was used, with a relatively slow pulsed beam excitation of 170 ns duration. Within our current interpretation, the elastic waves from Ref.~\cite{Pozar2018} associated to the optical force were therefore generated exclusively from the electrostriction effect. Although originating from a distinct optical force component, we expect that the same measurement scheme employing piezo-electric transducers reported in Ref.~\cite{Pozar2018} can be carried out for the setup simulated in Sec.~\ref{sec:disp} with $\tau=100$~fs and the associated displacements of 50 pm. Alternatively, modern interferometric techniques are capable of observing picometer-level deformations~\cite{Hu2019,Vine2009} and should provide a viable option as well.

The fast local heat deposition due to the optical absorption by the sample is very small, as the coefficient of optical absorption in fused silica is extremely low, reaching values of order $10^{-7}$~cm$^{-1}$~\cite{Loriette2003}, justifying our earlier assumption of a lossless medium. Nevertheless, elastic waves stemming from thermal stresses could still be generated and take part in the surface deformation~\cite{Nowacki1986}. In our model these waves were neglected, according to Eq.~(\ref{eq:TE}), as in fused silica glass the thermal expansion coefficient is also very low, approximately $5.5\cdot10^{-7}$~K$^{-1}$ at 25~$^{\circ}$C~\cite{Weber2003}. Thermal nonlinearities can also be typically neglected in sub-picosecond excitation regime~\cite{Boyd2020}. Therefore, the surface deformations in our system are expected to occur exclusively due to opto-mechanical effects induced by the optical force.

\section{Discussion}\label{sec:disc}
Observing the effects of the optical force in bulk material has been historically a challenging task due to its small magnitude and the general predominance of thermal effects~\cite{Borth1977}. Only recently quantitative works have been reported, in which the spatio-temporal behavior of the optical force density could be characterized in transparent liquids~\cite{Astrath2022,Astrath2023,Astrath2024}. However, because the pulsed excitation in these works is of order 1 ns, only the electrostriction effect was present -- thus, nothing about light's momentum could be inferred. In fact, the best theoretical model to describe momentum transfer in dielectric media currently is the so-called mass-polariton formulation~\cite{MP1,MP6,Anghinoni2024}, where the photon is coupled to a mass density wave within the material, which is driven forward by Abraham's force. In this formulation, the field has a momentum of Abraham's form (proportional to $n^{-1}$), while the mass-polariton quasi-particle has a momentum of Minkowski's form (proportional to $n$). The difference in these momenta is carried by the mass density wave, whose existence still lacks experimental confirmation but had already been theorized in earlier literature as a necessary material contribution in the field plus matter closed system~\cite{Jones1978b,Milonni2010,Loudon2005}. In any case, Abraham's optical force plays a central role in the mass-polariton theory, and thus its precise characterization in bulk matter is crucial to further test the validity of this model. 


Several quantitative experiments have been reported in the literature on the deformation of free liquid surfaces under illumination by laser beams~\cite{Astrath2014,Capeloto2015,Verma2015,Verma2017,Astrath2022,Verma2022,Verma2023,Verma2023b,Verma2024,Verma2024b,Zhang2015}. Some of these works attributed the observed deformation to light's momentum, but this poses an incorrect interpretation as even for pulsed beams we need, as shown in Sec.~\ref{sec:f}, short laser pulses with a few picoseconds duration for this contribution to be significant. None of the cited works meets this condition, as they employ either cw beams or pulsed beams with $\tau \gg 1$~ps -- therefore, the reported deformations must be solely due to stress effects occurring at the free liquid surface. Although not described in Eq.~(\ref{eq:f}), which is valid for homogeneous media, such stress at the surface can be shown to generate a radiation pressure as used for the boundary conditions in Sec.~\ref{sec:disp}. For a more detailed discussion on this topic and considerations on fluid dynamics, see Refs.~\cite{Brevik1979,Gordon1973,Lai1976,Leonhardt2014}.

The few available measurements of light's momentum reported a linear dependence on the medium's refractive index~\cite{Jones1954,Jones1978,Strait2019,Gibson1980,Campbell2005}. Although these works provide very valuable information on the overall optical momentum transfer, we notice that they are ultimately unable to give us details on how exactly the optical momentum force density behaves. This is so because their approach is focused in obtaining the optical momentum itself, which is related to the volume- and time-integrated optical momentum force. Quite generally, once a force density contribution is integrated to conform to a total force/center of mass dynamics description, we inevitably lose the more fundamental information on its spatiotemporal distribution~\cite{Silveirinha2017}. Such distribution can be relevant in certain applications -- especially the ones involving deformable matter samples~\cite{Torbati2022}. In any case, it is important to emphasize that the momentum component of the optical force density seen in Eq.~(\ref{eq:f}) is in complete agreement with light's momentum depending linearly on $n$. This can be formally shown by integrating the momentum force density in time and space while also carefully considering both field and material contributions, as shown analytically in Ref.~\cite{Milonni2010} for a plane wave and numerically in Refs.~\cite{MP1,Partanen2023,MP2} through simulations of opto-elastic dynamics with gaussian beam excitation.

Although the temporal dispersion was seen to produce no important changes in the beam propagation within the sample, we must note that the optical force given in Eq.~(\ref{eq:f}) is valid only for non-dispersive media. Fortunately, the Abraham force density in dispersive media is theoretically known (to first order in dispersion)~\cite{Partanen2021,Milonni2010}. Effectively, the factor $(n^2-1)$ in Eq.~(\ref{eq:f}) would be replaced by $(n n_\mathrm g-1)$, where $n_\mathrm g$ is the group refractive index, defined as $n_\mathrm g=c\beta_1$. As $n_\mathrm g$ is very often close in value to $n$ away from resonances, the dispersive Abraham force density is typically of the same order of magnitude of its non-dispersive version. Also, higher order contributions should be small, but could be added to the optical force if necessary by generalizing, for example, the derivation from Ref.~\cite{Milonni2010}. Therefore, with the argumentation presented we expect that the deformations seen in Fig.~(\ref{fig:comsol}) are not significantly altered by dispersion. A similar reasoning applies to nonlinear contributions, i.e., although Eq.~(\ref{eq:f}) is valid only for linear media, the nonlinear extra contributions that would occur in the optical force are not expected to influence the surface deformations significantly, as $n_2 I \ll n$.

With the detailed considerations presented, we argue that experimental systems similar to the one simulated in Sec.~\ref{sec:num} can be employed in seeking the first observation of the mechanical effects from the optical momentum force distribution. A full characterization of this force component can potentially provide novel functionalities to optomechanical devices and related applications as its behavior in space and time is distinct from its already known stress counterpart. Also, theoretical advances on fundamental physics are envisaged as the role of the optical momentum transfer in continuum matter would be further clarified. 

\section{Conclusions}\label{sec:conc}
We calculated the optical force generated in dielectric media due to the presence of short pulsed gaussian beams. The stress and the momentum contributions of the optical force were seen to present distinct behavior in both space in time, with the former acting in the transversal direction of beam propagation and the latter acting in the same direction of beam propagation. We showed that in typical optical setups pulses with duration of about 1~ps generate both contributions with the same order of magnitude. 

The optically-induced surface deformation of a fused silica glass sample was also simulated. For a pulsed beam with 100~fs duration, 100~$\mu$m radius and 1 mJ total energy, transient displacements of 50 pm were obtained. The properties of the chosen sample and characteristics of the optical setup allowed the pulse to propagate with no significant alterations in its shape due to the simultaneous action of nonlinear, dispersive and thermal effects, which greatly simplified the theoretical analyses. In this situation, the optical force density was dominated by the Abraham term, which corresponds unambiguously to a force component originating from the optical momentum transfer. Consequently, in our setup the surface deformations can be exclusively attributed to the momentum part of the optical force.

Displacements of picometric-order as obtained in our simulated setup can be measured by current piezo-electric sensors and interferometric techniques, and we briefly discussed how the former can be employed in our case. A successful measurement would provide pioneering characterization of the spatio-temporal distribution of the Abraham force density under optical regime, which would significantly clarify the ongoing Abraham-Minkowski fundamental problem on light's momentum transfer while also potentially contributing to a more versatile design of optomechanical-based applications.    

\section*{Acknowledgments}
The research leading to these results received funding from CNPq (304738/2019-0 and 307415/2022-8), CAPES (Finance Code 001), Funda\c{c}\~{a}o Arauc\'{a}ria, and FINEP. 

\appendix
\section{Calculation of the optical force}\label{app:f}
For a linearly polarized gaussian beam in fundamental mode propagating in the $z$-direction we have~\cite{Pedrotti1993}
\begin{equation}\label{eq:E}
    \bb E(\bb r,t) \!=\! E_0 \frac{w_0}{w(z)}\ee^{-\frac{r^2}{w^2(z)}}\ee^{-\frac{(t-\xi)^2}{\tau^2}}\ee^{i(\beta z-\omega t)}\hat{\bb x},
\end{equation}
where we assumed a gaussian time-dependence of the pulse and additional phase terms have been ignored as they are not important here. In Eq.~({\ref{eq:E}}), $E_0$ is the electric field amplitude, $w_0$ is the beam waist, $w(z)$ is the beam radius, $\xi$ is the time of maximum irradiance, $\tau$ is the temporal beam width, $i=\sqrt{-1}$, $\beta$ is the wavenumber, $\omega$ is the angular frequency and $\bb r=(r,\theta,z)$ is the position vector in cylindrical coordinates. The polarization direction is $\hat{\bb x}=\cos\theta\hat{\bb r}-\sin\theta\hat{\bm \theta}$. For a fixed total beam energy $Q$, the field amplitude $E_0$ scales as $E_0\propto\sqrt{(Q/w_0^2\tau)}$~\cite{Anghinoni2023b}.

Eq.~({\ref{eq:E}}) allows us to write the optical force density from Eq.~({\ref{eq:f}}) as
\begin{widetext}
\begin{equation}
    \bb f(r,z,t) = -4\left[\frac{r}{w^2(z)}\hat{\bb r}+\frac{2(t-\xi)}{c\tau^2} \hat{\bb z}\right]\varepsilon_0\frac{(n-1)}{(n+1)}\frac{w_0^2}{w^2(z)}E_0^2\ee^{-\frac{2r^2}{w^2(z)}}\ee^{-\frac{2(t-\xi)^2}{\tau^2}}.
\end{equation}
\end{widetext}
In the last equation, we have neglected the gradient in the $z$-direction as $\partial_z |\bb E|^2 \ll \partial_r |\bb E|^2$ in the paraxial approximation (or, equivalently, the Rayleigh range is much greater than the beam radius). The product $(\bb E \times \bb H)/2$, which corresponds to the beam intensity $I$, was approximated for air as $I(r,z,t) \approx \varepsilon_0 c |\bb E(r,z,t)|^2$/2. We also multiplied the amplitude $E_0$ by $2/(n+1)$ to account for the transmission under normal incidence for a beam incident from the air, as usual. The first term inside the brackets corresponds to the electrostriction and the second to Abraham's force density.

To estimate the relative magnitude of these two contributions of the optical force, we can take $w(z)\approx w_0$ in Eq.~(4), where $w_0$ denotes now an average beam radius inside the material of a well-collimated beam. It is also convenient to write the force density in terms of beam intensity instead of amplitude, yielding
\begin{equation}
    \bb f(r,z,t)\!=\! -\frac{8}{c}\left[\frac{r}{w_0^2}\hat{\bb r}\!+\!\frac{2(t\!-\!\xi)}{c\tau^2} \hat{\bb z}\right]\!\frac{(n\!-\!1)}{n\!+\!1}I(r,z,t), 
\end{equation}
which is exactly Eq.~(\ref{eq:f_I}).

The beam fluence, denoted by $F$, is defined as $F = \int I \dd t$. For a gaussian beam, we then have $F=\varepsilon_0 c \int |\bb E|^2 \dd t$. Performing this integration, it is easy to show that the peak value of $F$ is $2Q/(\pi w_0^2)$.

 \section{Ray transfer matrix analysis}\label{app:ray}
Consider Fig.~(\ref{fig:setup}). In our case, we want to find the distance $d$ the generates a beam radius of $w_0=100$~$\mu$m at the first sample surface and $w_L=114$~$\mu$m at the second sample surface, which are separated by a distance $L=1$~mm. This calculation must be performed for realistic values of the lens focal length $f$ and incoming collimated beam radius $w_\mathrm c$. This problem can be solved with the ray transfer matrix method (also known as ABCD matrix method), which corresponds to a very simple and versatile way to describe the propagation of a paraxial beam and its interaction with optical elements~\cite{Pedrotti1993}. Specifically, the propagation of the beam from plane 1 to plane 2 (which are arbitrary planes perpendicular to the optical axis) can be described by the complex radius of curvature of beam, $\tilde{q}$, as
\begin{eqnarray}\label{eq:q1}
    \tilde{q}_2 = \frac{A\tilde{q}_1+B}{C\tilde{q}_1+D},
\end{eqnarray}
where the coefficients $A,B,C,D$ represent the resultant effect of traversing the optical elements and propagating near the optical axis and are obtained as a multiplication of 2x2 matrices, where each matrix represents one of the optical events. Also, $\tilde{q}$ is defined as 
\begin{equation}\label{eq:q2}
 \frac{1}{\tilde{q}}=\frac{1}{R} + \frac{i\lambda}{\pi w^2},   
\end{equation}
 where $R$ is the (real) radius of curvature and $w$ is the beam waist. 

To obtain $d$, we must find the expressions for $\tilde{q}$ at both surfaces of the sample, where the beam radii are fixed. First, we must calculate $\tilde{q}$ at the focal spot. In ray transfer matrix context, this is obtained by a collimated incidence on a converging thin lens followed by the propagation to the focal spot. The coefficients are
\begin{equation}
 \begin{bmatrix}
A & B \\
C & D 
\end{bmatrix} 
=
 \begin{bmatrix}
1 & f \\
0 & 1 
\end{bmatrix} 
 \begin{bmatrix}
1 & 0 \\
-\frac{1}{f} & 1 
\end{bmatrix} 
 \begin{bmatrix}
1 & 0 \\
0 & 1 
\end{bmatrix} 
,
\end{equation}
which yields the complex radius at the focal spot $\tilde{q_f}=f^2/(\tilde{q}_\mathrm{c}+f)=-i\pi w_{\mathrm{min}}^2/\lambda_0$ , where $\tilde{q}_\mathrm{c}=-i\pi w_\mathrm{c}^2/\lambda_0$ is the complex radius of the incoming collimated beam and Eqs.~(\ref{eq:q1}) and~(\ref{eq:q2}) were used. From the $\tilde{q}_f$ equation we get $w_{\mathrm{min}}=[w_\mathrm c^{-2}+(\pi w_\mathrm c/\lambda_0 f)^2]^{-1/2}$, where we assumed $w_\mathrm c^2/\lambda \gg f$~\cite{Pedrotti1993}. 

Similarly, the complex radius at the first sample surface, $\tilde{q}_0$, is obtained by the subsequent propagation of $\tilde{q}_f$ by a distance $d$. This results in 
\begin{equation}
 \begin{bmatrix}
A & B \\
C & D 
\end{bmatrix} 
=
\begin{bmatrix}
1 & d \\
0 & 1 
\end{bmatrix} 
,
\end{equation}
which gives the complex radius at the first surface as $\tilde{q}_0 = \tilde{q}_f+d=(1/R_0+i\lambda_0/ \pi  w_0^2)^{-1}$.

To obtain the complex radius at the second sample surface, $\tilde{q}_L$, we add to $\tilde{q}_0$ the effects of the refraction at the flat interface followed by the propagation of distance $L$ inside the sample as
\begin{equation}
 \begin{bmatrix}
A & B \\
C & D 
\end{bmatrix} 
=
\begin{bmatrix}
1 & L \\
0 & 1 
\end{bmatrix} 
\begin{bmatrix}
1 & 0 \\
0 & \frac{1}{n}
\end{bmatrix} 
,
\end{equation}
which now gives $\tilde{q}_L = n\tilde{q}_0+L=(1/R_L+i\lambda_0/ \pi n w_L^2)^{-1}$.

Expanding and then separating the complex equations for $\tilde{q}_0$ and $\tilde{q}_L$ into real and imaginary parts, we get four equations -- one for $R_0$, one for $R_L$, one that involves $R_0$ and $d$ and one that involves $R_L$ and $d$. Notice that $d$ must satisfy both its equation simultaneously. Solving numerically for $d$ with $w_\mathrm c=1.5$~mm and $f=7.5$~cm, we obtain at last $d=5$~mm.

\bibliography{Refs}

\end{document}